\documentclass[a4paper,11pt]{article}
\usepackage{pos}
\usepackage{floatrow}

\title{Empirical assessment of cosmic ray propagation in magnetized molecular cloud complexes}
 \ShortTitle{CR propagation in molecular clouds}

\author*[a,b]{Ellis R. Owen}
\author[a,b]{Alvina Y. L. On}
\author[a]{Shih-Ping Lai}
\author[c,d]{Kinwah Wu}

\affiliation[a]{Institute of Astronomy, National Tsing Hua University, Hsinchu, Taiwan (ROC)}
\affiliation[b]{Center for Informatics and Computation in Astronomy, National Tsing Hua University, \\Hsinchu, Taiwan (ROC)}
\affiliation[c]{Mullard Space Science Laboratory, University College London, Holmbury St. Mary, \\Dorking, Surrey RH5 6NT, United Kingdom}
\affiliation[d]{Research Center for Astronomy, Astrophysics and Astrophotonics, Macquarie University, \\Sydney, NSW 2109, Australia}

\emailAdd{erowen@gapp.nthu.edu.tw}

\abstract{Molecular clouds are complex magnetized structures, with variations over a broad range of length scales. Ionization in dense, shielded clumps and cores of molecular clouds is thought to be caused by charged cosmic rays (CRs). These CRs can also contribute to heating the gas deep within molecular clouds, and their effect can be substantial in environments where CRs are abundant. CRs propagate predominantly by diffusion in media with disordered magnetic fields. 
The complex magnetic structures in molecular clouds therefore determine the propagation and spatial distribution of CRs within them, 
and hence regulate their local ionization and heating patterns.
\vspace{0.2cm}

Optical and near-infrared (NIR) polarization of starlight through molecular clouds is often used to trace magnetic fields. The coefficients of CR diffusion in magnetized molecular cloud complexes 
can be inferred from the observed fluctuations in these optical/NIR starlight polarisations. Here, we present calculations of the expected CR heating patterns 
in the star-forming filaments of IC 5146, determined from optical/NIR observations. 
Our calculations show that local conditions give rise to substantial variation in CR propagation. This affects the local CR heating power. 
Such effects are expected to be severe in star-forming galaxies rich in CRs. 
The molecular clouds in these galaxies could evolve differently 
to those in galaxies where CRs are less abundant.}

\FullConference{37$^{\rm{th}}$ International Cosmic Ray Conference (ICRC 2021)\\
		July 12th -- 23rd, 2021\\
		Online -- Berlin, Germany}

\begin{document}
\maketitle

\section{Introduction}

\noindent 
The interstellar medium (ISM) of 
galaxies   
is multi-phase, 
with complex hydrodynamics and thermal structures 
spanning a broad range of scales.   
Cold, dense neutral gas regions containing molecular clouds (MCs) and filaments are intermingled with hot, tenuous gases 
in (approximate) pressure equilibrium.  
These are permeated by magnetic fields with 
  complex configurations and multiple length-scales.     
Radiation from stars maintain a high ionization fraction throughout the hot tenuous material comprising most of the ISM volume. 
In the dense cores of MC complexes, 
 densities are sufficiently high to shield material from much of the ionizing (particularly ultraviolet, UV) interstellar radiation 
 with dust and molecular Hydrogen~\citep{Draine2011Book}.   
However, observations have shown 
  sustained ionization 
  of rates of up to $\zeta^{\rm H} = 10^{-17} - 10^{-15}~\text{s}^{-1}$ in the dense cores of MCs in our Galaxy  
  \cite[e.g.][]{Tak2000}, 
  and 
  $\zeta^{\rm H} = 10^{-16}~\text{s}^{-1}$ 
  in diffuse 
  interstellar clouds~\cite[e.g.][]{Black1978ApJ}. 

The cause of this ionization is widely attributed to cosmic rays (CRs)~\cite{Goldsmith1978}, which would also act to regulate the gas temperature~\citep[e.g.][]{Spitzer1968ApJ}. 
In this work, we consider an empirical propagation and heating model of CRs in MC environments, which was first introduced in~\cite{Owen2021ApJ}. This accounts for CR cooling and interaction mechanisms relevant to CR species (protons and electrons) in a self-consistent manner. We apply this model to the polarization observations of MCs and their filaments within the Milky Way, from which CR propagation parameters and heating patterns can be determined. We then demonstrate that, if a MC complex with a similar configuration to those found in the Milky Way, were irradiated by a flux of CRs similar to those inferred for nearby starburst galaxies, the CR heating effect could be important enough to substantially raise the Jeans mass within its densest regions. This would have important and widespread implications for star-formation, where a different mode of star-formation with a higher stellar initial mass function and/or delayed onset could be maintained in galaxies or regions harboring a large CR reservoir.

\section{Method}

\subsection{CR propagation in MC complexes}
\label{sec:prop_theory}

\noindent
The propagation of CRs can be described using the transport equation:
\begin{align}
      \frac{\partial n}{\partial t} -
      \nabla \cdot \left[D(E, {\boldsymbol{s}})\nabla n \right] 
      + \nabla \cdot \left[{\boldsymbol{v}} n\right]+ \frac{\partial}{\partial E} \left[ \;\! b(E, {\boldsymbol{s}}) n \;\! \right]   =  Q(E, {\boldsymbol{s}}) - S(E, {\boldsymbol{s}}) \ ,  
\label{eq:transport_equation}     
\end{align}
\citep[e.g.][]{Schlickeiser2002_book}. Here, $n = n(E, {\boldsymbol{s}})$ is the differential number density of CRs (number of CR particles per unit volume per energy interval between $E$ and $E+{\rm d}E$) 
  at a location 
  ${\boldsymbol{s}}$. The `source' and `sink' terms ($Q$ and $S$, respectively) encode the injection and absorption of CRs into a system. The influx of particles may be specified by a boundary condition, at which the irradiating CR spectrum is defined. The choice of spectral model is discussed in detail in~\cite{Owen2021ApJ}, however its exact form (if reasonable) does not substantially affect the results of this work. 
   The diffusive term
  $\nabla \cdot \left[D(E, {\boldsymbol{s}})\nabla n \right]$ is governed by the coefficient
  $D(E, {\boldsymbol{s}})$
  which depends on the gyro-scattering radius (or frequency)
  of the CRs of energy $E$ in their local magnetic field, 
  as well as
  turbulence and 
  magnetohydrodynamical (MHD) perturbations along local magnetic field vectors. This is dependent on the detailed magnetic field configuration of a system (see section~\ref{sec:empirical_determination}). 
  The second propagation term $\nabla \cdot \left[{\boldsymbol{v}} n\right]$ is usually an advection term which describes the propagation of CRs in the bulk flow of a magnetized medium. When applying equation~\ref{eq:transport_equation} to a MC, however, it can be used to describe the propagation of CRs through a magnetized ISM. Due to the CR streaming instability, 
  the effective CR `flow' velocity typically corresponds to the Alfv\'{e}n 
  speed~\cite[e.g.][]{Commercon2019AA}, 
  $v_{\rm A}$.\footnote{This approximation is not universal. It would break down for  
  particles with energies above $\sim$ 10 GeV, as they do not experience 
  sufficient scattering in the ISM~\cite{Chernyshov2018NPPP}.}
  
It is useful to decompose CR transport into two hierarchical regimes, one at the `cloud-scale', where macroscopic transport behaviors emerge, and one on more local scales where the full diffusion prescription in the local magnetic field is required.
CRs are constrained to gyrate around and propagate along the magnetic field lines. Their flux, and any change in their number density  
 on the `cloud' scale, are scaled with 
  the number density of magnetic field lines at a location. 
Hence, we may consider a parametrisation 
   in terms of the magnetic-field concentration:  
   $\chi (\equiv {B}/{B_0})$ 
   \citep[see][]{Desch2004ApJ}, 
   which is the ratio of the magnetic field strength $B$ measured 
  at a location within the MC compared to the magnetic field strength external to the cloud, $B_0$. 
This follows from the magnetic field strength 
  being defined as the magnetic flux density through a point. 
  If the magnetic fields through a MC have a macroscopic, inward-curved (`hourglass') morphology, this can act to both focus the CRs through the cloud according to the shape of the field, as well as mirror them away from the densest regions due to conservation of kinetic energy and magnetic moment (a process called magnetic mirroring). 
  This can be quantified by $\eta(\chi) = 4\pi \left[\;\! \chi - \sqrt{\chi^2-\chi}\;\!\right]$~\cite{Desch2004ApJ}, which we adopt as a large-scale adjustment to the solution of equation~\ref{eq:transport_equation} so that only localized values for the CR transport parameters through a MC (in particular, the diffusion coefficient $D$) are necessary. While this is an approximation, and a more thorough treatment would consider the variation of $\eta$ through the cloud by solving the transport equation directly, we consider it to be sufficient for our purposes and better facilitates application to observational data.

\subsection{Empirical diffusion parameter} 
\label{sec:empirical_determination}

\noindent
The diffusion parameter for CRs is determined by the strength and structure of local magnetic fields. Charged CRs scatter more strongly in magnetic fields where perturbation length-scales are comparable to the gyro-radius of the CR. If perturbations are much smaller than this, they do little to modify the propagation of CR. If they are much larger, they would guide the large-scale propagation of an ensemble of particles (cf. the mirroring process described in section~\ref{sec:prop_theory}).
The spatial CR diffusion coefficient $D$ may be related to the structure of the magnetic field through the Fokker-Planck (FP) coefficients.
 As the flow of CRs 
and the orientation of the magnetic fluctuations are largely parallel to the orientation of the background cloud-scale magnetic field vector, the only non-vanishing FP coefficient is
\begin{equation}
    P_{\mu \mu} \approx \frac{\mathcal{J}(\lambda_1)}{\;\!v_{\rm A}\;\! \lambda_1}\left(\frac{\omega_{\rm L}\;\!B_0}{B}\right)^2\;\!\mathcal{I}_{\perp} \ .  
    \label{eq:pmumu_original}
\end{equation}
Here $\lambda_1 =\lambda_{\rm td}\;\!(|\omega_{\rm L}|/\omega_{\rm p,0})$ 
  is the CR resonant scattering length scale 
  parallel to the background magnetic field line.  
The turbulent decay length scale is
  $\lambda_{\rm td} \approx v_{\rm A}\;\!\tau_{\rm td}$, 
  (where we use $\tau_{\rm td} =2$\,Myr -- for discussion, see~\cite{Owen2021ApJ}).
The CR gyro-frequency is
    $\omega_{\rm L}$, 
with a sign convention set by the charge (in units of proton charge).
The normalization
 $\omega_{\rm p,0}$ is taken to be
 the gyro-frequency of a CR proton $\omega_{\rm p}$ at a reference energy of 100 MeV.
We adopt a magnetic field strength normalization
$B_0 = 1\,{\rm m}\text{G}$ 
to be comparable to the magnetic field strength outside the densest parts of clumps/cores. $\mathcal{I}_{\perp} = \pi/k_{\rm c}^2$ specifies the contribution from orthogonal components of the magnetic field fluctuations, where $k_{\rm c}$ is the wavenumber normalization, defined as $\omega_{\rm p,0}/v_{\rm A}$.
  The dimensionless variable $\mathcal{J}(\lambda_1)$ 
  encodes the magnetic field structure in terms of its fluctuations 
  along the direction of the background large-scale magnetic field vector. It is defined as 
\begin{equation}
    \mathcal{J}(\lambda_1) \equiv
    \int_0^{\lambda_1} {\rm d}\lambda \ \frac{ \lambda}{\lambda_1} 
     \hat{P}_\parallel(k_{\rm c}\lambda) 
    + \int_{\lambda_1}^{\infty} {\rm d}\lambda\ \frac{\lambda_1}{\lambda}\;\! 
    \hat{P}_\parallel(k_{\rm c}\lambda) \ ,  
\end{equation} 
   where $\hat{P}_\parallel$ is the power spectrum of the fluctuations 
  along the large-scale magnetic field vector.
The diffusion coefficient now follows as:
\begin{equation}
    D \approx \frac{c^2}{8}\left(1-\frac{1}{\gamma^2}\right)\int_{-1}^{1} {\rm d}\mu \;\! \frac{(1-\mu^2)^2}{P_{\mu \mu}} \ .
    \label{eq:diff_coeff_full}
\end{equation}
\cite{Owen2021ApJ}, where $\gamma$ is the CR Lorentz factor, $c$ is the speed of light, and $\mu$ is the cosine of the CR pitch angle. 
Since the pitch angles are small, 
  $P_{\mu \mu}$ 
  is not strongly dependent {on} $\mu$ \citep[see][]{SA1993_2}, 
  such that 
 $D \approx {2\;\!c^2}/{15\;\!P_{\mu \mu}}$\footnote{Assuming the MC is in a steady-state.},  
 for $\gamma\gg 1$. 
 
 To empirically determine the diffusion coefficient $D$, a discretized form of the fluctuation statistic $\mathcal{J}$ is required. This is given by:
\begin{equation}
    \mathcal{J}(\lambda_1) \approx
    k_c^{-1}\;\!\sum_{n=1}^{i_{\rm b}} \frac{\kappa_{n}}{k_c \lambda_1}\;\! P(\kappa_n) + 
    k_c^{-1}\;\!\sum_{n=i_{\rm b}}^{N_{\rm bins}} \frac{k_c \lambda_1}{\kappa_{n}}\;\! P(\kappa_n) \ ,
    \label{eq:j_n1_discrete}
\end{equation}
which is applied to a discrete data-set separated into $N_{\rm bins}$ bins according to scale. In this study, we bin the measured polarization angle (PA) difference between pairs of points according to the angular separation of each pair. The discrete Fourier Transform (FT) of the angular dispersion function $S_2(\ell_n)$ can be taken for each scale-bin $\ell_n$ to find $P(\kappa_n)$, according to the relationship between the power spectrum $\hat{P}(k)$ and angular dispersion function $\mathcal{S}_2(\ell)$ (cf. the Wiener-Khinchin theorem  \cite{Wiener1930_book, Percival1995_book}, which gives  $\hat{P}(k) = \frac{1}{2}\mathcal{F}\left[\mathcal{S}_2(\ell)\right]$, 
where ${\mathcal F}[...]$ denotes a FT). 
$i_{\rm b}$ is the bin index 
corresponding to the (normalized) resonant length scale $\lambda_1\;\!k_c$, and $\kappa_n$ is the normalized wavenumber associated with $\ell_n$. 

\section{Applications}

\subsection{IC 5146}
\label{sec:ic51436region}

\begin{figure}
\floatbox[{\capbeside\thisfloatsetup{capbesideposition={right,center},capbesidewidth=6cm}}]{figure}[\FBwidth]
{\caption{Dispersion functions calculated for the 4 bands, from \cite{Owen2021ApJ}.
    The increasing power at larger scales due to the curved hour-glass magnetic field structure
    is not shown here as it is not relevant to our fluctuation analysis. 
    $x$ error bars indicate the bin size. $y$ error bars are 1$\sigma$ Gaussian errors estimated by a Monte Carlo approach with 10,000 perturbations. }\label{fig:sf_profile}}
{\includegraphics[width=6.7cm]{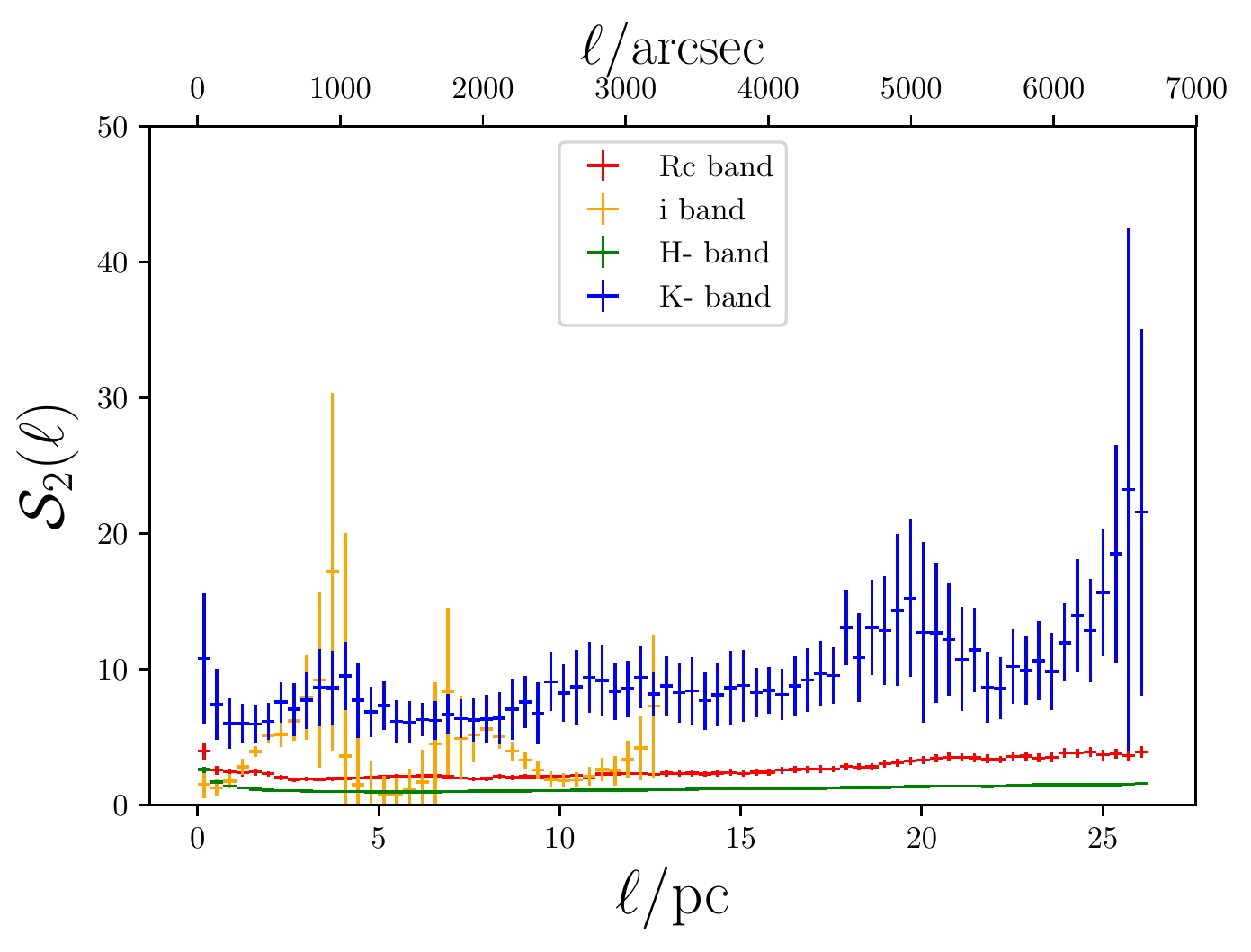}}
\end{figure}

\noindent
IC 5146 is an interstellar MC complex, located in Cygnus. 
It exhibits a converging filamentary system of dark clouds, and elongated sub-structures extending from a main filament (as studied with \textit{Herschel} observations~\cite{Arzoumanian2011}). 
The formation of the system is thought to be driven by large-scale turbulence ~\cite{Arzoumanian2011}. This would introduce perturbations into the otherwise well-ordered large-scale magnetic field morphology~\cite{Wang2017ApJ, Wang2019ApJ}, making it an ideal demonstration test-case for our empirical diffusion parameter estimation method. 

For our calculations, we assume IC 5146 to be a single interstellar cloud located at a distance of $813\pm106~\text{pc}$~\cite{Dzib2018ApJ}. We use optical and near infra-red stellar polarization observations in four bands ($R_{\rm c}$, $i$, $H$ and $K$) towards IC 5146 \citep{Wang2017ApJ} to trace the magnetic fields and their fluctuations, which we use to empirically determine the diffusion coefficient of CRs through the region. In every band, we compute the angular dispersion function using a bin size of 90'' (0.35 pc), which gives reasonable signal-to-noise ratios at the scales of interest (see Figure~\ref{fig:sf_profile}). The discrete FT of the dispersion function is then computed, with the diffusion coefficient for the full IC 5146 region following from equation~\ref{eq:diff_coeff_full} (using the magnetic field strengths for this cloud computed in~\cite{Wang2020ApJ}). We do not find any clear support for variation of the diffusion parameter in different regions of the cloud, and values determined from each observational band are roughly consistent with one another~\citep{Owen2021ApJ}.

The empirically-determined diffusion coefficient is then used to solve equation~\ref{eq:transport_equation} for 15 of the distinct filamentary structures identified in IC 5146~\cite{Arzoumanian2011} for which sufficient parametric information is available (column density $N_{\rm H}$ and inner length scale $R_{\rm flat}$). The resulting total CR heating rate and equilibrium temperatures (when balancing \textit{only} CR heating against cooling) in the filament ridges, i.e. their densest point where the effects of CRs are likely to be most substantial, is shown in Table~\ref{tab:filaments}. This demonstrates that stronger heating typically occurs 
  in the filaments that have larger volume densities. However, the effect is not directly proportional to the density (cf. filament 6 vs 7). This is because of the corresponding increase in magnetic field strength within a filament, which increases the importance of magnetic mirroring and deflection. While these processes operate antagonistically, the relative importance of each is determined by the spatial extent of the cloud and the variation of the density (and, hence, magnetic field) profile, where less steep density (and magnetic) profiles lead to less mirroring. This increases the CR heating efficiency through a filament. The strongest heating arises in filaments 6 and 12, which have a high density but a relatively shallow density profile. This configuration reduces magnetic mirroring/deflection effects and allows a higher flux of CRs to penetrate into the filament. Despite the action of CR heating, the equilibrium temperatures it can sustain are far lower than those expected for the dense filaments of MCs. This would suggest that CR activity alone cannot maintain substantial core/filament temperatures in this system. Instead, other processes (e.g. heating from mechanical processes or turbulence dissipation) may be more important in maintaining clump/filament temperatures at the expected level of around $\sim 10 \;\! {\rm K}$ under Galactic conditions. 

\begin{table*}
\resizebox{0.7\textwidth}{!}{%
\centering
\begin{tabular}{|l|c|c|c|c|c|c|}
\hline
  ID & $p$ & $R_{\rm flat}\;\!/\;\!\text{pc}$ & $n_{\rm c}\;\!/\;\!10^4\;\!\text{cm}^{-3}$ & $\mathcal{H}\;\!/10^{-26}\;\!\text{erg}\;\!\text{cm}^{-3}\;\!\text{s}^{-1}$ & $T_{\rm eq, CR}\;\!/\;\!~\text{K}$ \\
 \hline  
 \hline 
 \!\!
 1   & 2.1 & 0.09 & 0.3 & 0.59 & $0.8$ \\ 
 2   & 1.9 & 0.1 & 0.7 & 9.2 & $1.7$ \\ 
 4   & 1.4  & 0.04 & 0.7 & 4.3 & $1.3$ \\
 5   & 1.5  & 0.02 & 7 & 68 & $2.5$ \\
 6   & 1.7  & 0.07 & 4 & 290 & $4.0$ \\
 7   & 1.6  & 0.05 & 2 & 33 & $2.2$ \\
 8   & 1.5  & 0.09 & 0.4 & 6.8 & $1.8$ \\
 9   & 1.5  & 0.07 & 0.8 & 16 & $2.0$ \\
 10   & 2.1  & 0.1 & 0.5 & 2.5 & $1.2$ \\
 11   & 1.9  & 0.07 & 1 & 6.5 & $1.5$ \\
 12   & 1.5  & 0.05 & 4 & 240 & $3.7$ \\
 13   & 1.6  & 0.04 & 3 & 4.3 & $2.3$ \\
 20   &  1.5 & 0.05 & 0.2 & 0.34 & $0.7$ \\
 21   &  1.7 & 0.09 & 0.3 & 1.9 & $1.2$ \\
 25   & 1.5  & 0.05 & 0.7 & 4.9 & $1.4$ \\
\hline
\end{tabular}}
\caption{Specific rates of CR heating $\mathcal{H}$ and ionization $\zeta^{\rm H}$
  in 15 of the 27 filaments   
  of the IC 5146 region identified by \protect\cite{Arzoumanian2011}, where 
  filament ID numbers here correspond to those used in that paper.
} 
\label{tab:filaments}
\end{table*}

\subsection{Nearby starburst galaxies}

\begin{figure*}
\includegraphics[width=0.9\textwidth]{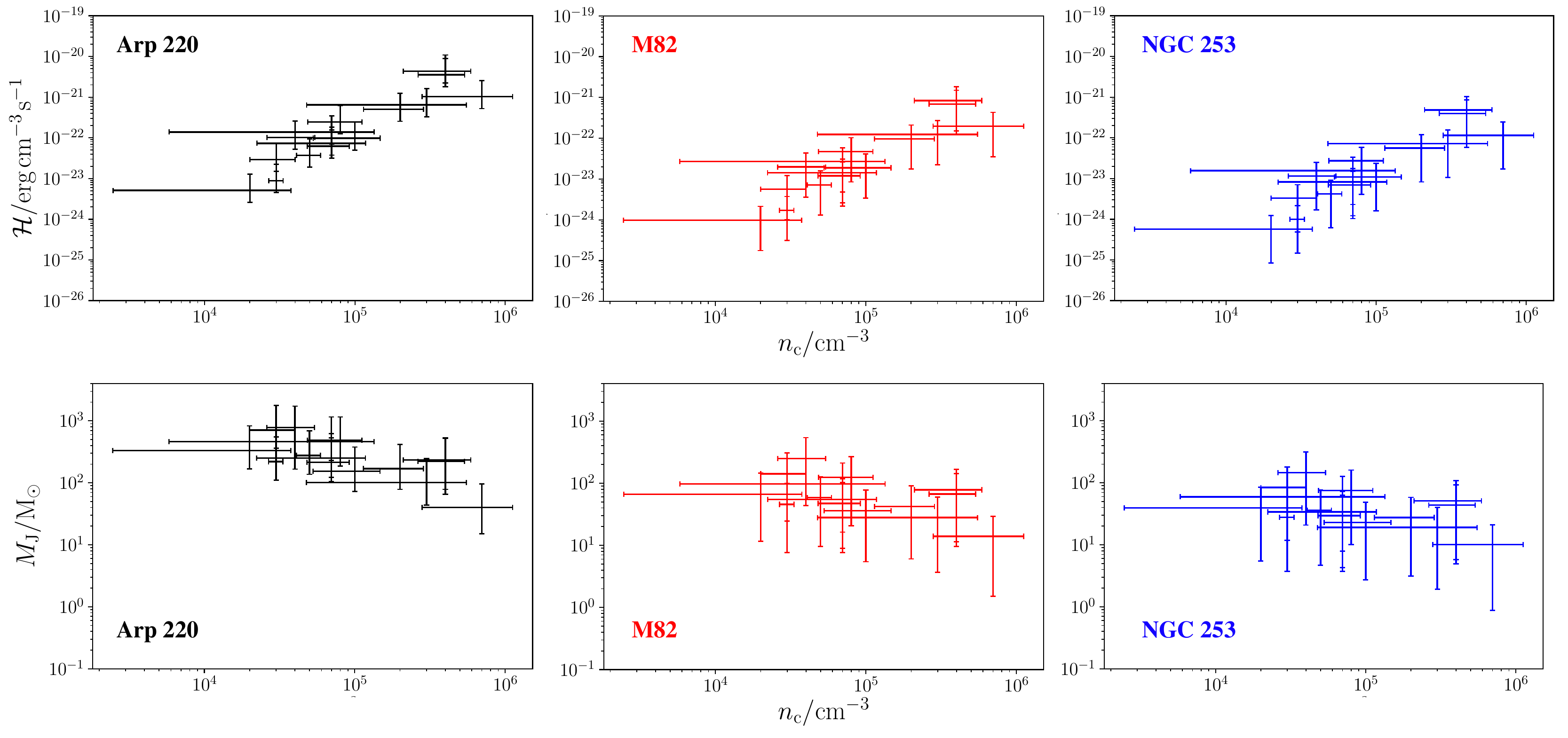}
\caption{\textbf{Top row}: Heating power by CRs if the filaments of IC 5146 (Table~\ref{tab:filaments}) were irradiated by a flux of CRs as that in Arp 220, M82 or NGC 253. \textbf{Bottom row}: Corresponding Jeans' masses for the same filaments, if assuming heating through the clouds is dominated by CRs. $y$-axis uncertainties reflect the possible range of CR energy densities in each galaxy (as estimated in~\cite{Yoast-Hull2015MNRAS, Yoast-Hull2016MNRAS}), $x$-axis uncertainties follow from~\cite{Arzoumanian2011}. }\label{fig:starbursts_heating}
\end{figure*}

\noindent
Section~\ref{sec:ic51436region} demonstrated that, while CRs can have some heating impact in MC filaments within the Milky Way, it is not likely to dominate in typical Galactic ISM MC complexes. However, under conditions where the level of CR irradiation is higher, the impact of CR heating could be more substantial. Here, we apply our model to the same filamentary structures considered in section~\ref{sec:ic51436region}, but consider levels of CR irradiation comparable to the ISM of starburst galaxies. To estimate the CR flux, we consider CR energy densities determined for three nearby starbursts, NGC 253, M82 and Arp 220. The CR heating power that would be felt in the MC filaments is shown in the top row of Figure~\ref{fig:starbursts_heating}, plotted against their peak density, where uncertainties in heating power follow from the range of possible CR energy densities in each of the three starburst galaxies considered~\cite{Yoast-Hull2015MNRAS, Yoast-Hull2016MNRAS}. The heating impact of CRs in these three environments is clearly strengthened compared to Galactic ISM conditions. Moreover, the heating power is correlated with the energy density of CRs within each system - Arp 220 being the highest and NGC 253 being the lowest of the three starbursts considered. A correlation between heating power and gas volume density is also evident in all cases, and this correspondence is not sensitive to the level of CR irradiation.

The implications of this heating could be significant. 
The Jeans mass sets the level of thermal pressure against gravitational collapse, and is determined by temperature and gas volume density by $M_{\rm J} \propto T^{3/2} \;\! n^{-1/2}$. 
The corresponding Jeans' masses for each of the filaments when irradiated by the three starburst CR fluxes is shown in the lower row of Figure~\ref{fig:starbursts_heating}. Even though temperature increases due to CR heating in some of the filaments in this case are modest, the impact on the Jeans' mass is substantial. 
While this is a crude measure of the maximum stable mass against collapse, it can still give an idea of the size of MCs and resulting stars/stellar clusters. 
Without CR heating, Galactic clouds could reach around $200~\text{M}_{\odot}$ before becoming unstable to gravitational collapse. 
However, with Arp 220 CR energy densities, this could increase by more than an order of magnitude, suggesting a delayed onset of star-formation in filaments strongly irradiated by CRs, or even the emergence of a different, \textit{top-heavy} mode of star-formation in CR-abundant environments~\cite[cf.][]{Papadopoulos2010ApJ}.

\section{Summary}

\noindent
We demonstrate how the propagation of CRs can be determined empirically from optical and near-infrared polarization of starlight through MC complexes within our Galaxy. This allows their impacts in the intricate, magnetized filaments of MC complexes to be determined in a more self-consistent manner, through regions where the magnetic field structure can be inferred. We show that the feedback impact (via heating) is unlikely to be substantial in Galactic environments, if the irradiating CR flux is comparable to the interstellar mean. However, the mass distribution of clouds/clumps in systems with higher CR energy densities (e.g. in star-forming galaxies) can become distorted. This may have important implications for ongoing star-formation in environments rich in CRs, where delayed or top-heavy star-formation may arise.

\acknowledgments

\noindent
ERO and AYLO are supported by the Ministry of Education of Taiwan at the Center for Informatics and Computation in Astronomy, National Tsing Hua University. SPL acknowledges a grant from the Ministry of
Science and Technology of Taiwan 109-2112-M-007-010-MY3.

\bibliographystyle{ICRC}
\bibliography{references}

\end{document}